# Essential role of long non-coding RNAs in de novo chromatin modifications: The genomic address code hypothesis

Ken Nishikawa[1] and Akira R. Kinjo[2,*]

[1]National Institute of Genetics, Research Organization of Information and Systems, Mishima, Shizuoka, 411-8540, Japan; [2]Institute for Protein Research, Osaka University, Suita, Osaka, 565-0871, Japan.

*Correspondence to ARK (akinjo@protein.osaka-u.ac.jp).

## Abstract

The epigenome, i.e. the whole of chromatin modifications, is transferred from mother to daughter cells during cell differentiation. When de novo chromatin modifications (establishment or erasure of, respectively, new or pre-existing DNA methylations and/or histone modifications) are made in a daughter cell, however, it has a different epigenome than its mother cell. Although de novo chromatin modifications are an important event that comprises elementary processes of cell differentiation, its molecular mechanism remains poorly understood. We argue in this Letter that a key to solving this problem lies in understanding the role of long non-coding RNAs (lncRNAs)- a type of RNA that is becoming increasingly prominent in epigenetic studies. Many studies show that lncRNAs form ribonucleo-protein complexes in the nucleus and are involved in chromatin modifications. However, chromatin-modifying enzymes lack the information about genomic positions on which they act. It is known, on the other hand, that a single-stranded RNA in general can bind to a double-stranded DNA to form a triple helix. If each lncRNA forms a ribonucleo-protein complex with chromatin-modifying enzymes on one hand and, at the same time, a triple helix with a genomic region based on its specific nucleotide sequence on the other hand, it can induce de novo chromatin modifications at specific sites. Thus, the great variety of lncRNAs can be explained by the requirement for the diversity of "genomic address codes" specific to their cognate genomic regions where de novo chromatin modifications take place.

## Introduction

While all cells constituting a multicellular organism are derived from a single zygote to share the identical genome, they have different epigenomes depending on their cell types. The epigenome is a genome-wide pattern of gene-expression regulation embodied as chromatin modifications composed of DNA methylation as well as histone post-translational modifications, such as acetylation, methylation, phosphorylation and so on. The epigenome is maintained through cell division via epigenetic memory transfer from mother to daughter cells. This process is well illustrated by the maintenance of methylated DNA through DNA replication, where hemi-methylated nascent DNA strands are selectively methylated with DNA methyltransferase DNMT1 to reproduce the original methylated DNA (Alberts et al. 2015).

Alteration of the epigenome, on the other hand, should occur in accordance with the progression of ontogenesis of an organism via cell differentiation. De novo chromatin modifications, including DNA methylations and histone modifications that are newly introduced or deleted, can be regarded as elementary steps of the epigenomic alteration. To date, the molecular mechanism of de novo chromatin modification is only poorly understood (Baubec et al. 2015). The difficulty of the problem may arise from the reciprocal relationship between the genome and the epigenome (Fig. 1). The downward blue arrows pointing from the epigenome to the genome in Fig. 1 represent the regulation of genomic information (gene expression), the molecular mechanisms of which are well understood. On the contrary, the upward red arrows pointing from the genome to the epigenome represent establishment or alteration of the epigenome (chromatin modifications) according to the information encoded on the genome possibly in conjunction with environmental cues (Fig. 1), the molecular mechanism of which is still unclear. Cell differentiation is a typical epigenetic phenomenon, during the course of which the epigenome is to be altered and a new epigenome specific to the terminally differentiated cell must be established. De novo chromatin modification may be a good starting point to tackle this basic biological issue.

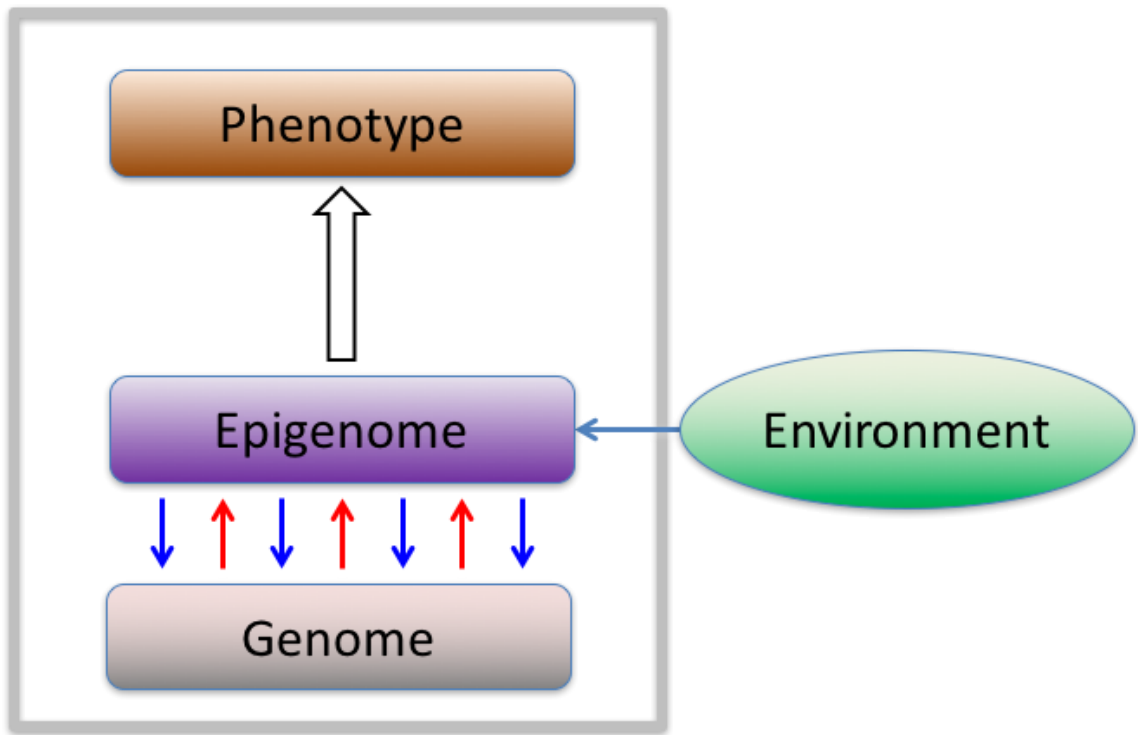

Figure 1. A schematic view of genome, epigenome and phenotype in a cell. The genome, under a certain environment (blue horizontal arrow), determines the epigenome (upward red arrows) while the epigenome regulates the genome by activating or repressing gene expressions (downward blue arrows). The genome covered with a specific set of epigenomic marks determines the phenotype (the upward white arrow).
______________________________________________________________________

**Cell differentiation and de novo chromatin modifications**

Suppose that new chromatin modifications that were absent in a mother cell are induced in the daughter cells over the course of cell differentiation. In the case of DNA methylation, for example, these correspond to de novo DNA methylations, that is, new methyl groups are added to previously unmethylated genomic sites. Similarly, it is expected that new histone modifications be introduced as cell differentiation proceeds.

Alteration of genome-wide chromatin modifications associated with embryogenesis as well as ontogenesis has been investigated by advanced sequencing techniques. Bisulphite sequencing, for instance, has helped to reveal the DNA methylation landscapes of mammals and plants. In mammals, methylation reprogramming over all the genomic regions, except for imprinted genes, are performed twice in primordial germ cells and immediately after fertilization (Reik et al. 2001). After this the DNA methylation restarts around the time of implantation, and DNA methylations as well as histone modifications rapidly proceed throughout the embryogenesis (Vastenhouw and Schier 2012). This process must involve massive de novo chromatin modifications.

It was observed in mammals that the DNA methylation globally covers the genome, including intergenic DNA regions as well as gene-bodies, leaving only CpG islands mainly localized in gene promoters and cis-regulating enhancers unmethylated (Suzuki and Bird 2008). Promoter and/or enhancer DNA regions are differentially methylated depending on different cell lineages and developmental stages. The differential methylation along the course of cell differentiation must be brought about by de novo DNA methylation. There are, however, few investigations to elucidate the molecular details of de novo chromatin modifications.

In the case of new chromatin modifications such as de novo DNA methylations, we cannot assume that the modifications are copied from a mother cell to daughter cells. We propose that the key to understanding the de novo chromatin modification, or the problem of how new epigenomes are established (the upward red arrows in Fig. 1), resides in non-coding RNAs, among them, long non-coding RNAs in particular, that are massively transcribed from the genome, as recent experiments have elucidated. Non-coding RNAs are encoded in the genome and expressed as transcripts, which matches with the direction of the upward red arrows in Fig. 1.

**Long non-coding RNAs are abundant**

In the early 21st century, the RIKEN Mouse Genome Project Team (Hayashizaki and Kanamori 2004), while analyzing cDNAs to identify genes in the mouse genome, discovered the existence of a wide variety of transcripts that did not correspond to any (protein-coding) genes (Katayama et al. 2005). Closer examination revealed that some of these transcripts overlapped with protein coding regions but were coded on the anti-sense strand while others are coded in intergenic regions. In short, as confirmed by later studies, the transcripts were found to originate from various genomic regions. Further studies have shown that a major fraction of the entire genome is transcribed and the numerous transcripts with unknown functions are called "dark matter RNAs" (Kapranov and St Laurent 2012).

These transcripts are now called "non-protein coding RNAs" (ncRNAs) as they do not code proteins. Depending on their length, they are roughly classified into small non-coding RNAs (snRNA) consisting of 20-40 nucleotides or long non-conding RNAs (lncRNAs) of more than 200 nucleotides. Here, we focus on lncRNAs. Similarly to ordinary mRNAs, lncRNAs are transcribed by RNA polymerase II (Pol II), and the transcripts are processed with 5' capping, splicing, and 3' polyadenylation. However, lncRNAs do not contain ORFs and are not translated to proteins. Compared to mRNAs, lncRNAs are generally less conserved, which makes it difficult to predict their functions by sequence homology. In addition, they are highly tissue-specific or cell-type specific and many of them have a low expression level.

The number of identified lncRNA sequences encoded in human genomes has increased each year. This curious phenomenon is related to the tissue-specific nature of lncRNAs. A definitive method to identify genes is to express the genes in cells and to analyze the resulting cDNAs. This procedure is especially essential for identifying such "ambiguous" genes as those encoding lncRNAs. The high tissue specificity implies, however, that only a limited variety of lncRNAs are expressed in a particular cell. It follows that we need to study many kinds of tissues and cells to obtain a comprehensive list of lncRNAs. The number of lncRNA-coding genes in the human genome was once estimated to be around 15,000 (Derrien et al. 2012), which was less than that of the protein-coding genes (about 21,000). However, the number has gradually increased since then and we now (in 2015) know at least 58,000 lncRNA-coding genes (Iyer et al. 2015). Iyer et al. (2015) have used many kinds of cancer cells to identify the lncRNAs (presumably because they could not use normal human cells for the investigation). These findings naturally raise the following two questions. (1) What are the roles of so many lncRNAs in the cells? (2) Why are so many lncRNAs necessary?

**LncRNAs are involved in various biological activities**

The transcripts originating from about two thirds of the genomic DNAs of human and other mammals were called the "dark matter", and researchers have wondered if they are of any biological significance or simply meaningless products transcribed from junk DNAs. Nevertheless, it has been reported that the number of lncRNAs correlates with evolutionary complexity of organisms better than the genome size or the number of protein-coding genes (Kapusta and Feschotte 2014). This observation does suggest some biological significance of lncRNAs.

In fact, as studies of epigenetics advanced, lncRNAs were found to be involved in various important biochemical, cellular and developmental activities. Some well-known examples include such lncRNAs as *Xist* and *Tsix* that are involved in the inactivation of X chromosomes, or as *Kcnq1ot1*, *Airn*, and *H19* in genome imprinting. The HOX gene cluster, a developmental-control DNA region important in embryogenesis, encodes the lncRNAs *HOTTIP* and *HOTAIR* that regulate the expression of HOXA genes and HOXD gene, respectively. More than 200 lncRNAs including *lnc-RoR* are known to be involved in the maintenance of the pluripotency of ES cells and/or iPS cells. LncRNAs also play vital roles in ontogenesis of tissues and organs and cell differentiation. Some examples are the following: Differentiation of the fat cell (*ADINR* and other several hundred lncRNAs); In the hematopoietic lineage, red blood cell differentiation (*lnc-EPS* and more than 400 lncRNAs) and T-cell differentiation (*lnc-MAF4* and more than 100 lncRNAs); Development of the heart (e.g., *Fendrr* and *Braveheart*); Development of the brain and neurons (*Evf2*, *RMST*, *Paupar*, *TUG1* and many other lncRNAs) (Lopez-Pajares 2016).

Many lncRNAs are expressed in various tissues and organs and are involved in their development and differentiation. Embryogenesis, ontogenesis and cell differentiation are accompanied with the establishment and alternation of epigenomes. Thus, it is natural to expect that lncRNAs play some role in the establishment and alteration of epigenomes (the upward red arrows in Fig. 1). This expectation was confirmed by recent studies using the epigenomic footprinting technique: the experiment was carried out for about 100 distinct cell types to indicate the concordance of cell-type specific transcription of lncRNAs with cell-type specific histone modifications (Amin et al. 2015). Now the question is: what are the precise roles of lncRNAs in epigenomic modifications?

To identify the function of lncRNAs, it is necessary to conduct experiments for each molecular species, which is in general more difficult than functional characterization of proteins. Thus, the number of functionally characterized lncRNAs is not large (about several hundreds). Nevertheless, we can already see a great diversity of their functions. One functional classification is to divide into two classes depending on whether they function inside the nucleus or in the cytoplasm. Examples of those lncRNAs functioning in the nucleus include those involved in chromatin modifications. Those functioning in the cytoplasm include those anti-sense lncRNAs that hybridize with their mRNA counterparts to inhibit the translation. Another classification is whether they are cis- or trans-regulatory. A lncRNA is said to be "cis-regulatory" if it functions in a genomic region near the coding region of the lncRNA itself. Otherwise, a lncRNA is said to be "trans-regulatory". While most of lncRNAs are thought to be cis-regulatory, some examples of trans-regulatory lncRNAs are known. One famous example is the lncRNA *HOTAIR* which is encoded in one of Homeobox genes, HOXC gene cluster on human chromosome 12. *HOTAIR* represses the expression of the HOXD gene on human chromosome 2. Thus, *HOTAIR* clearly acts in-trans (Fatica and Bozzoni 2014).

**LncRNAs have two functional domains**

Looking across lncRNAs with known functions, we notice that many of them form a ribonucleo-protein complex. In the following, we focus on the cases where the protein components are chromatin-modifying enzymes. Accordingly, the corresponding lncRNAs function in the nucleus.

One of the best characterized lncRNA-binding proteins is the PRC2 (Polycomb repressive complex 2), a chromatin-modifying (histone-methylation) complex consisting of several proteins (Geisler and Paro 2015). PRC2 binds a lncRNA by recognizing its stem-loop secondary structure. The specificity of the RNA-protein binding is low in the following sense. Since any sufficiently long RNAs tend to contain some stem-loop secondary structures, PRC2 almost indiscriminately binds a wide range of RNAs to form a ribonucleo-protein complex. This promiscuous RNA binding ability of PRC2 (Davidovich et al. 2013) is an important factor that resolves the mystery of the asymmetry between the limited number of chromatin-modifying enzymes and the large variety of lncRNAs.

LncRNAs bind not only to proteins, but also to DNAs or other RNAs. A single-stranded RNA can hybridize with another single-stranded DNA or RNA. It is also known that a single-stranded RNA can bind to a double-stranded DNA to form a triple-stranded helix (Buske et al. 2011; Li Y et al. 2016). The hybridization of an RNA and DNA is supposedly highly specific as it is based on complementary base pairs. Thus, a lncRNA can find DNA regions complementary to its DNA binding region to form a RNA-DNA helix. A longer binding region can achieve both higher affinity and higher specificity.

This picture of lncRNAs is in accordance with a previously proposed model in which lncRNAs have two functional domains (Johnson and Guigó 2014). According to this model, one functional domain of a lncRNA forms a stem-loop secondary structure which binds to a protein, and the other domain binds to the genomic DNA to form a triple helix. The two functional domains have distinctly different binding properties: the binding specificity is low in the former (RNA-protein) and high in the latter (RNA-DNA). That is, a particular protein can bind many different lncRNAs while a particular lncRNA can bind to only one (or few) specific DNA region(s). As already noted above, PRC2 can bind many lncRNAs by recognizing a stem-loop secondary structure. TrxG (Trithorax Group Complex) that has an "opposite" function to PRC2 (TrxG *activates* gene expression by introducing a histone modification H3K4me3 while PRC *represses* gene expression by depositing H3K27me3) also binds lncRNAs in a similar manner (Fatica and Bozzoni 2014).

Today, several lncRNAs are known to form both chromatin-modifying ribonucleo-protein complex and RNA-DNA triple helix, which we review next.

**LncRNAs anchor chromatin modifiers to genomic DNA sites**

Schmitz et al. (2010) studied de novo DNA methylation of rDNA (ribosomal RNA-coding gene) promoter in mouse. They found that a kind of non-coding RNA (ncRNA), called promoter RNA (pRNA), is involved in the DNA methylation. pRNA is about 200-nt long and encoded on the rDNA promoter region. When transcribed, pRNA forms a triple helix about 20-nt long between a complementary pRNA sequence and double stranded DNA of the rDNA promoter region. The triple helix is recognized by de novo DNA methyltransferase DNMT3b for binding. DNMT3b, anchored to the genomic DNA via the pRNA, deposits the methyl group on a single cytosine of CpG sequence in the rDNA promoter, thereby hindering the binding of transcription factor TTF-1 to the promoter, which inactivates the transcription of the rDNA gene. Furthermore, the central region of the pRNA makes a stem-loop secondary structure which is recognized by a chromatin remodeling protein complex called NoRC. NoRC introduces histone modifications (H4K20me3 and H3K27me3) to the nearby chromatin for heterochromatin formation, which, together with the DNA methylation, strongly represses the rDNA gene (Schmitz et al. 2010). This study showed that a single non-coding RNA, namely a pRNA, mediates not only de novo DNA methylations but also de novo histone modifications.

Other examples are also provided in a recent review by Li Y et al. (2016) who discussed the detailed mechanisms of triple helix formation by lncRNAs and DNAs in vivo. As they have already pointed out, six out of seven lncRNAs that they listed are reported to form triple helices as well as to recruit some chromatin modifiers (see Table 1 in the paper by Li Y et al. 2016). We would like to point out that these proteins are also known to be involved in de novo chromatin modifications. The list includes, in addition to the pRNA discussed above, *Fendrr* (Foxf1 and Pitx2 genes and PRC2 and TrxG/MLL complexes), *Khps1* (SPHK1 gene and the histone acetyltransferase p300/CBP), *PARTICLE* (MAT2A gene and PRC2), *MEG3* (TGF-β pathway genes and PRC2) and *HOTAIR* (PCDH7, HOXB2 and other genes, and PRC2 and the histone demethylase LSD1) (Li Y et al. 2016). The same, or similar, mechanism of de novo chromatin modifications is expected for other pRNAs as well as eRNAs (enhancer RNAs) that have been recently observed in various gene promoters and enhancers (Li W et al. 2016).

**Each lncRNA harbors the genomic address code of its own?**

Chromatin-modifying enzymes do not have the positional information of the genomic sites on which they act. We propose that it is the role of lncRNAs to provide the missing positional information to the chromatin-modifying enzymes. Since this information specifies positions on the genome, we would like to call it "genomic address code (GAC)", rather than "cellular address code" (Batista and Chang 2013). Several concrete examples were described in the previous section. We predict the existence of GAC regions to be ubiquitous in those lncRNAs that bind to de novo chromatin-modifying enzymes such as PRC2 and TrxG. Although there are only a handful experimental studies to date confirming the existence of GACs in lncRNAs, if it is indeed the case, we can clearly understand the role of lncRNAs in epigenomic regulation. That is, a lncRNA binds a chromatin-modifying enzyme by using its stem-loop and anchors it to a particular site of the genomic DNA specified by its GAC by forming a triple helix, and the enzyme then modifies the chromatin. If this hypothesis holds, it is possible for chromatin-modifying complexes to be recruited to arbitrary genomic sites simply by modifying the information of the GAC sequence in lncRNAs. This mechanism provides a simple way to increase the complexity of gene expression patterns by increasing the variety of lncRNAs, which may account for the correlation between the number of lncRNAs and evolutionary complexity of organisms (Kapusta and Feschotte 2014). We can thus explain the molecular mechanism of de novo chromatin modifications and understand not only that lncRNAs are indispensable factors in the process of establishing and altering epigenomes (the upward red arrows in Fig. 1), but also why tens of thousands of lncRNAs are required for determining the epigenome in various types of cells.